\documentclass{ptapap}
\usepackage{amsmath}
\usepackage{amssymb}

\newenvironment{myfont}{\fontfamily{lmtt}\selectfont}{\par}
\DeclareTextFontCommand{\textmyfont}{\myfont}

\author{Swayamtrupta Panda}[CFT,LNA,CAMK]
\author{Ewa Julia Skorek}[UU]
\affil[CFT]{Center for Theoretical Physics, Polish Academy of Sciences, Al. Lotnik\' ow 32/46, 02--668 Warsaw, Poland}
\affil[CAMK]{Nicolaus Copernicus Astronomical Center, Polish Academy of Sciences, Bartycka 18, 00--716 Warsaw, Poland}
\affil[LNA]{Laborat\'orio Nacional de Astrof\'isica - MCTIC, R. dos Estados Unidos, 154 - Na\c{c}\~oes, Itajub\'a - MG, 37504--364, Brazil}
\affil[UU]{Warsaw University Observatory, Al. Ujazdowskie 4, 00--478 Warszawa, Poland}

\title{Metallicity evolution in quasars}

\begin{document}

\maketitle

\begin{abstract}
Broad-band spectra of active galaxies contain a wide range of information that help reveal the nature and activity of the central continuum source and their immediate surroundings. Understanding the evolution of metals in the spectra of Active Galactic Nuclei (AGN) and linking them with the various fundamental black hole (BH) parameters, for example, BH mass, the bolometric luminosity of the source, its accreting power, can help address the connection between the growth of the BH across cosmic time. We investigate the role of selected metallicity indicators utilizing the rich spectroscopic database of emission lines covering a wide range in redshift in the recent spectroscopic data release of Sloan Digital Sky Survey (SDSS). We make careful filtering of the parent sample to prepare a pair of high-quality, redshift-dependent sub-samples and present the first results of the analysis here. To validate our findings from the simple correlations, we execute and evaluate the performance of a linear dimensionality reduction technique - principal component analysis (PCA), over our sub-samples and present the projection maps highlighting the primary drivers of the observed correlations. The projection maps also allow us to isolate peculiar sources of potential interest.
\end{abstract}

\section{Introduction}
The chemical properties of galaxies have been considered as one of the key parameters in understanding galaxy evolution, one that is closely related to the star formation history \citep{Maiolino_2019}. Active galaxies, owing to their large luminosity range, can be observed over a broad range of redshift. An important constituent of the AGN are the broad-line region (BLR) that are primarily responsible for the broad emission lines observed in a typical AGN spectrum. A handful of diagnostic ratios in the UV and optical have been tested to study and constrain the metal content in the BLR, e.g., (i) N V$\lambda$1240/C IV$\lambda$1549; (ii) Fe II(2700-2900\AA)/Mg II$\lambda$2800; hereafter Fe II(UV)/Mg II; and (iii) Fe II(4434-4684\AA)/H$\beta\lambda$4861; hereafter Fe II(opt)/H$\beta$ \citep{hamannferland92,panda19b,Shin_2021}. These diagnostic ratios probe the metal content of BLR at different redshift ranges and a possible link between them, if established, can allow to gain further insights in the AGN-host co-evolution across cosmic time.

\section{Metallicity indicators and correlations}

In order to evaluate the efficacy of the aforementioned metallicity indicators that span a wide range in the redshift and keeping in mind the spectral coverage of SDSS (3600 - 10,400 \AA), we prepare a pair of high-quality sub-samples using the DR14 quasar catalogues \citep{rakshitetal2019}: (A) a low-redshift sample (0.617 $\leq z \leq$ 0.89) containing 6,072 sources that encompass the Fe II(UV), Mg II, Fe II(opt) and H$\beta$ emission lines; and (B) a high-redshift sample (2.09 $\leq z \leq$ 2.25) containing 1,950 sources that cover the N V, C IV, Fe II(UV) and Mg II emission lines.

\begin{figure}[!h]
    \centering
    \includegraphics[width=0.85\columnwidth]{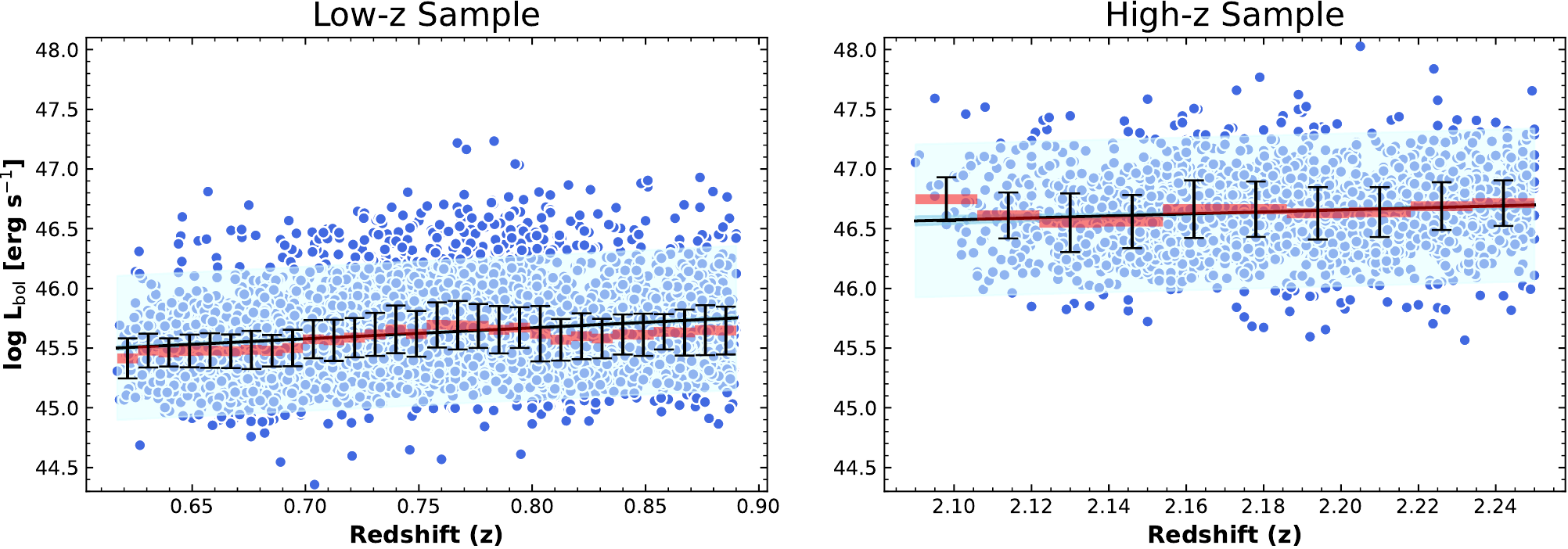}\\
    \includegraphics[width=0.85\columnwidth]{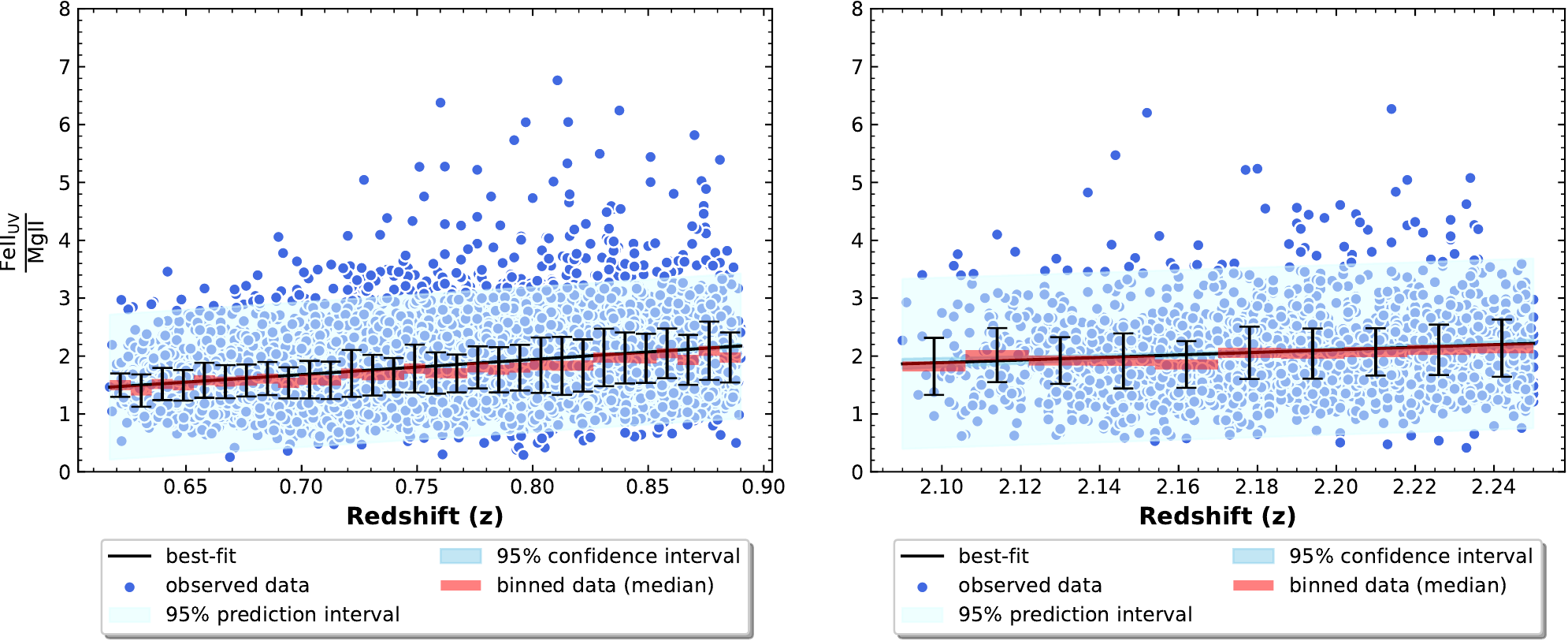}
    \caption{Upper panels show the bolometric luminosity distribution as a function of redshift for the (LEFT) low-z, (RIGHT) high-z sub-samples. We median-bin the data-sets accounting for the relative number of sources in each sub-sample. A linear-fit with 95\% confidence and prediction intervals are also shown. Lower panels show the corresponding behaviour for the ratio Fe II(UV)/Mg II as a function of redshift.}
    \label{fig1}
\end{figure}

A physical connection between the Fe II(UV)/Mg II ratio and BLR metallicity can be understood in terms of metal enrichment from star formation. As a proxy of the Fe/Mg abundance ratio, we can track the change of Fe II(UV)/Mg II as a function of time from the onset of star formation based on an enrichment delay \citep[tens of Myr to a few Gyrs,][]{Matteucci_2001} between $\alpha$-elements (e.g., Mg) and iron which are produced mainly from Type II and Type Ia supernovae, respectively \citep[see][for an overview]{Shin_2021}. The correlation between several line flux ratios, tracing the chemical properties of the BLR, and Eddington ratio can be considered as the relation between the BLR metallicity and accretion activity of AGN. We consider here the Fe II(UV)/Mg II ratio but will extend the study to include other indicators in our analysis. 

From Figure \ref{fig1} (upper panels) we can clearly notice the change in the gradient of the distribution - going from a steeper slope ($\sim$0.92, in the left panel) to a shallower one ($\sim$0.84, in the right panel), in addition to the rise in the net luminosity. These are consistent with the conclusions from prior works involving the study of evolution of quasar luminosity function and its similarity with the progression of star-formation rate density across cosmic time. We confirm a similar behaviour for the ratio Fe II(UV)/Mg II (slope decreases from 2.61 to 2.22, see bottom panels in Figure \ref{fig1}).

\section{What PCA reveals for our samples?}

\begin{figure}
    \centering
    \includegraphics[width=0.45\columnwidth]{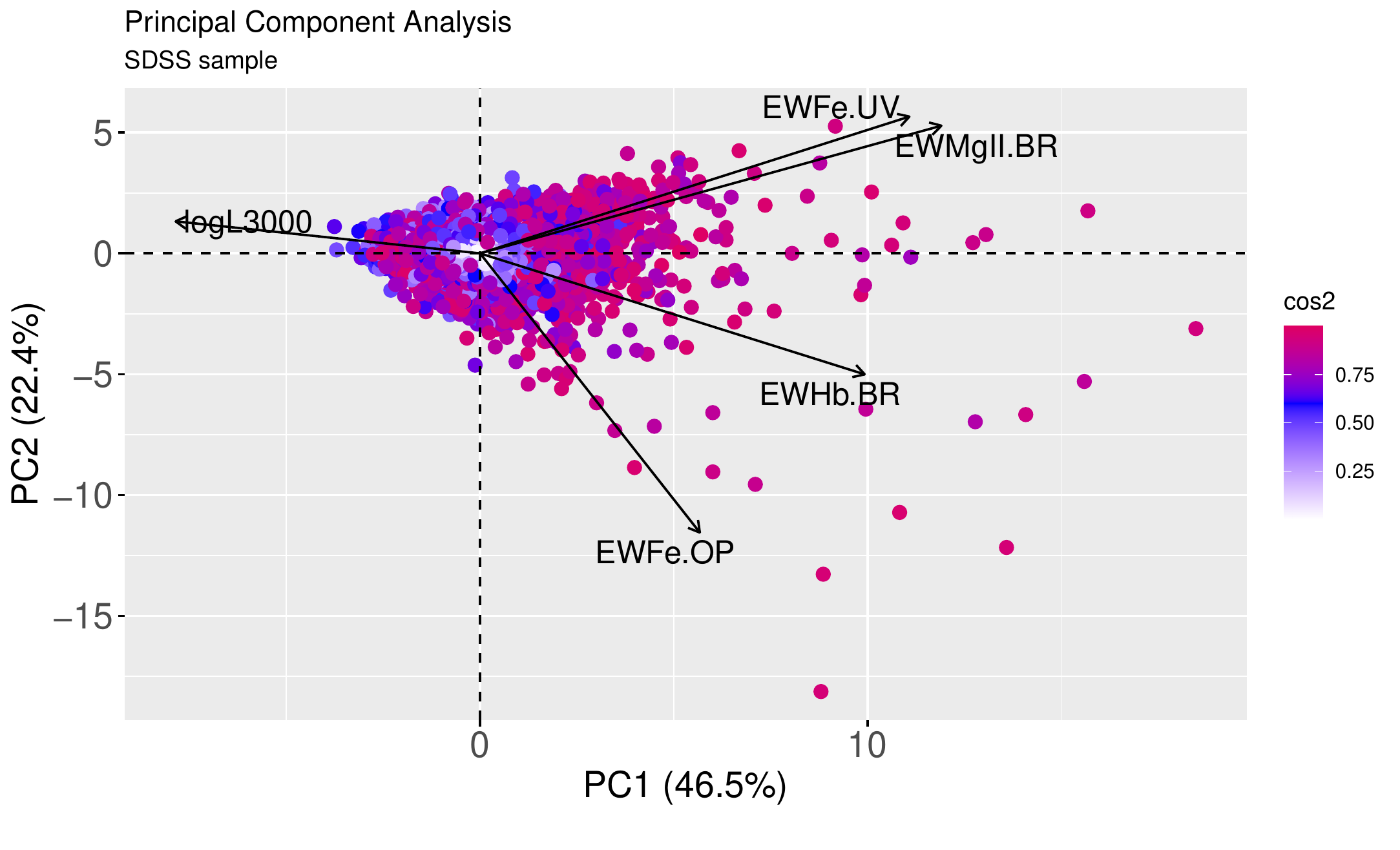}
    \includegraphics[width=0.45\columnwidth]{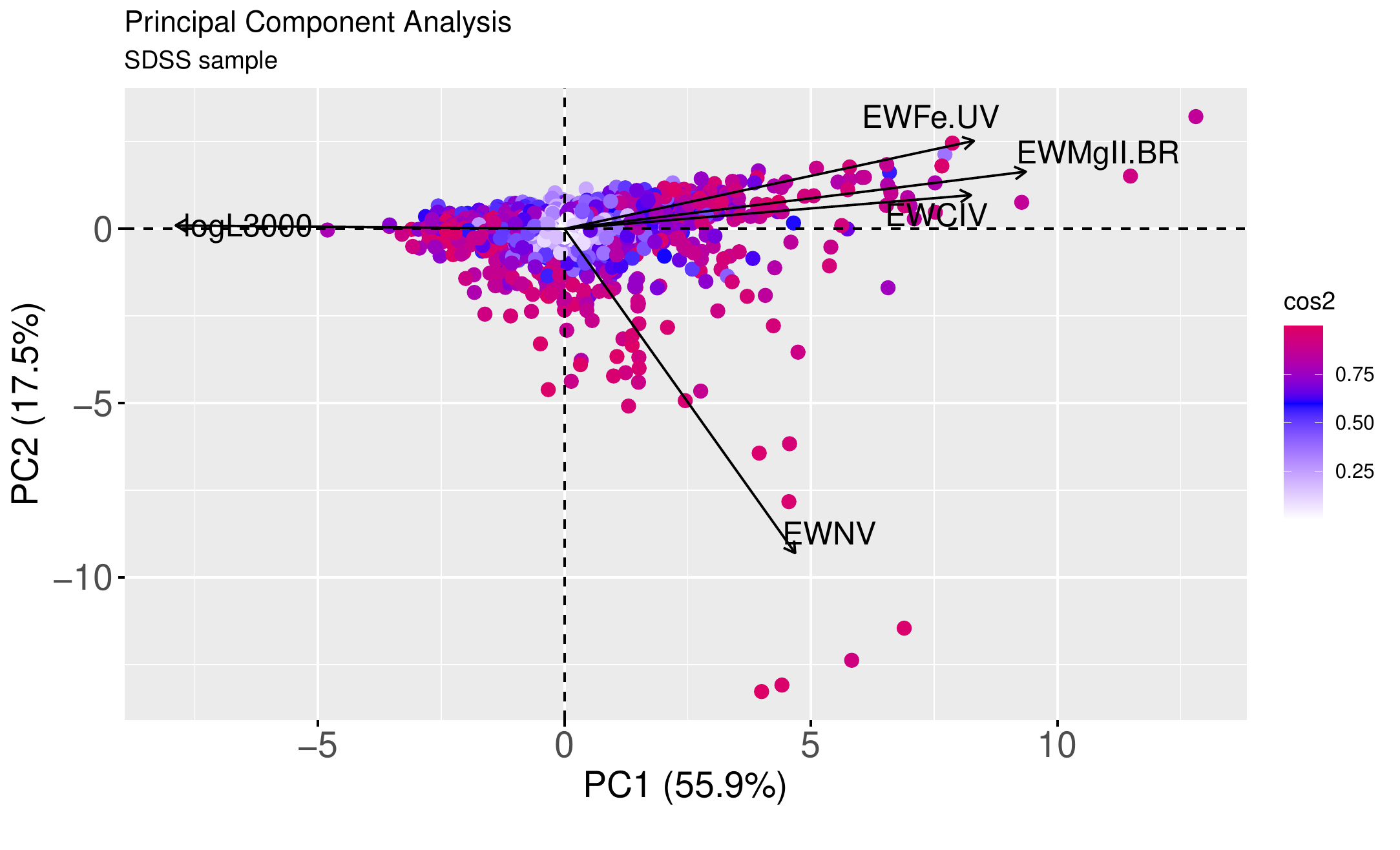}\\
    \caption{Correlations between the first two principal components (PC1 and PC2) for the (LEFT) low-z sub-sample, and (RIGHT) high-z sub-sample. The contribution of the respective principal components to the overall variance in the data-sets (\textmyfont{cos2}) is also noted.}
    \label{fig2}
\end{figure}

After standardizing our data-sets, we run a PCA utilizing five observables, each for the two sub-samples, obtained directly from the spectral fitting of the SDSS spectra. The correlation space between the first two PCs are shown in Figure \ref{fig2} for each sub-sample. The similarity in the projection maps indicate the continuity in the sub-samples across redshift in addition to the closeness between the line widths of Fe II(UV) and Mg II.

\section{Highlights and future work}
Through this pilot study:
(a) we re-confirm the usability of Fe II(UV)/Mg II ratio as a robust indicator of the quasar evolution across cosmic time. Our preliminary PCA results have supplemented the conclusions obtained from direct correlations. PCA has also allowed us to gather peculiar sources. The analysis of the spectra of these sources in ongoing, (b) we plan to experiment with other linear and non-linear clustering algorithms and benchmark their performances, and, address the physical mechanisms at play using radiative transfer modelling accounting for the change in the ionizing continuum across a wide range of redshift.

\acknowledgements{The project was partially supported by the Polish Funding Agency National Science Centre, project 2017/26/\-A/ST9/\-00756 (MAESTRO  9), MNiSW grant DIR/WK/2018/12 and acknowledges partial support from CNPq Fellowship (164753/2020-6).}

\bibliographystyle{ptapap}
\bibliography{panda2}

\begin{thebibliography}{6}
\providecommand{\natexlab}[1]{#1}
\providecommand{\url}[1]{\texttt{#1}}
\providecommand{\urlprefix}{URL }
\providecommand{\eprint}[2][]{\url{#2}}

\bibitem[{{Hamann} \& {Ferland}(1992)}]{hamannferland92}
{Hamann}, F., {Ferland}, G., \emph{The Astrophysical Journal Letters}
  \textbf{391}, L53 (1992)

\bibitem[{{Maiolino} \& {Mannucci}(2019)}]{Maiolino_2019}
{Maiolino}, R., {Mannucci}, F., \emph{Astronomy \& Astrophysics Reviews}
  \textbf{27}, 1, 3 (2019)

\bibitem[{{Matteucci} \& {Recchi}(2001)}]{Matteucci_2001}
{Matteucci}, F., {Recchi}, S., \emph{The Astrophysical Journal} \textbf{558},
  1, 351 (2001)

\bibitem[{{Panda} et~al.(2019){Panda}, {Marziani}, \& {Czerny}}]{panda19b}
{Panda}, S., {Marziani}, P., {Czerny}, B., \emph{The Astrophysical Journal}
  \textbf{882}, 2, 79 (2019)

\bibitem[{{Rakshit} et~al.(2020){Rakshit}, {Stalin}, \&
  {Kotilainen}}]{rakshitetal2019}
{Rakshit}, S., {Stalin}, C.~S., {Kotilainen}, J., \emph{The Astrophysical
  Journal Supplement Series} \textbf{249}, 1, 17 (2020)

\bibitem[{{Shin} et~al.(2021)}]{Shin_2021}
{Shin}, J., et~al., \emph{The Astrophysical Journal} \textbf{917}, 2, 107
  (2021)

\end{thebibliography}

\end{document}